\documentclass[12pt,preprint]{revtex4}
\usepackage{graphicx} 
\usepackage{amsmath}
\usepackage{amsfonts,amssymb}
\usepackage[matrix,arrow,curve,frame,poly,arc]{xy}
\usepackage[english]{babel}

\def\be{\begin{equation}}
\def\ee{\end{equation}}
\def\ba{\begin{eqnarray}}
\def\ea{\end{eqnarray}}

\begin{document}
\bibliographystyle{plainnat}

\title{Painlev\'e analysis for two 1D parabolic-parabolic models of chemotaxis; some travelling wave solutions}

\author{Maria Shubina }

\email{yurova-m@rambler.ru}

\affiliation{Skobeltsyn Institute of Nuclear Physics\\Lomonosov Moscow State University
\\ Leninskie gory, GSP-1, Moscow 119991, Russian Federation}


\begin{abstract}

In this paper we study the Painlev\'e analysis for two models of chemotaxis. 
We find that in some cases the reductions of these models in terms of travelling wave variable allow exact analytical solutions. 

\end{abstract}

\keywords{Painlev\'e analysis, Patlak-Keller-Segel model, parabolic-parabolic system, soliton solution}

\maketitle

\section{Introduction}
\label{intro}

Chemotaxis, or the directed movement of cells (bacteria or other organisms) up or down a chemical concentration gradient, plays an important role in many biological and medical fields such as embryogenesis, immunology, cancer growth. The macroscopic classical model of chemotaxis was proposed by Patlak in 1953 \cite{P} and by Keller and Segel in 1970s \cite{KS1}-\cite{KS3}. This model describes the space-time evolution of the cells density $ u(t,x) $ and the concentration of the chemical substance $ v(t,x) $. The general form of this model is:
\be \left\{
\begin{aligned}
u_{t}-\nabla (\delta_{1} \nabla u - \eta_{1} u \nabla \phi (v)) & =0 \\ 
v_{t}-\delta_{2} \nabla^{2} v - f(u, v) & =0, \nonumber
\end{aligned}
\right.
\ee
where $ \delta_{1} > 0 $ and $ \delta_{2} \geq 0  $ are the cells and chemical substance diffusion coefficients respectively, $ \eta_{1} $ is the chemotaxis coefficient (when $ \eta_{1} > 0 $ this is an attractive chemotaxis, or "positive taxis", and when $ \eta_{1} < 0 $ this is a repulsive, or  "negative" one \cite{Ni}, \cite{Li&Wang}). The functions $ \phi (v) $ is the chemosensitivity function and $ f(u, v) $ characterizes the chemical growth and degradation. The chemosensitivity function is taken in some forms, particularly in linear $ \phi (v) \sim v $ and logarithmic $ \phi (v) \sim  \ln v $ ones. The model with linear $ \phi $ often is considered with $ f(u, v) = \tilde{\sigma} u - \tilde{\beta} v $, or simpler $ f(u, v) = u - v $, and it is studied in a large number of paper. The case of positive $  \tilde{\sigma} $ and nonnegative $ \tilde{\beta} $ is studied in \cite{J&L}, \cite{N&S&Y}, \cite{C&E&M}, \cite{T&W}, \cite{F}. However it is necessary to point out that the sign of $ \tilde{\sigma} $ may effect on the mathematical properties of the system. So, $ \tilde{\sigma} = 1 $ means an increase of a chemical substance, proportional to cells density, whereas $ \tilde{\sigma}= - 1 $ -- its decrease, what changes the solvability conditions of the system \cite{TF}. The review article \cite{H1} summarizes the different mathematical results. 

The model with logarithmic $ \phi (v)$ is actively studied too. For the case of $f(u, v) = - v^{m} u +  \tilde{\beta} v $ the extensive analysis is performed in \cite{Wang}. This survey is focused on different aspects of traveling waves solutions. When $ \tilde{\beta} = m = 0 $ the traveling waves were considered in \cite{N&I}; when $ \tilde{\beta} = 0 $ and $ m = 1 $ the system was studied in \cite{Rosen}. The existence of global solutions is established in \citep{W}. 

In this paper we consider two one-dimensional simplified models with the coefficients $ \delta_{1} $, $ \delta_{2} $ and $ \eta_{1} $ are constants.  In the first model the chemosensitivity function is taken linear $ \phi (v) = v $ and $ f(u, v)= \tilde{\sigma} u - \tilde{\beta} v  $: 
\be \left\{
\begin{aligned}
u_{t}- u_{xx} + (u v_{x})_{x} & =0 \\ 
v_{t}-\alpha v_{xx} + \beta v - \sigma u & =0.\\ 
\end{aligned}
\right.
\ee
The second system to be discussed has $ \phi (v) = \ln v $ and  $ f(u, v)= \tilde{\beta} v - \tilde{\sigma} v u $, $ \tilde{\beta}, \tilde{\sigma} $ are constants:
\be 
\label{eq:2} 
\left\{
\begin{aligned}
u_{t}- u_{xx} + \eta (u \frac{v_{x}}{v})_{x} & =0 \\ 
v_{t}-\alpha v_{xx} - \beta v + \sigma v u & =0,\\ 
\end{aligned}
\right.
\ee
where the coefficients are the ones remaining after the replacement $ t \rightarrow  \delta_{1} t$ (and $ v \rightarrow \frac{\eta_{1}}{\delta_{1}} v $, $ u \rightarrow \frac{\eta_{1}}{\delta_{1}} u $ for (1)), $ \sigma = \pm 1 $, $ x \in \Re, t \geq 0 $, $ u=u(x,t) $, $ v=v(x,t) $. 

We study the Painlev\'e analysis for the systems (1) and (2). Using the results of this analysis we solve the systems under consideration in terms of travelling wave variables. Unfortunately, not all solutions can have biological interpretation since $ u $ and $ v $ become negative for some domain of variable. However we believe that they are interesting as examples of exact solitary (soliton-like) solutions for chemotaxis models.

\section{Painlev\'e analysis end exact solutions}
\label{sec:1}

Let us study the Painlev\'e analysis for the systems (1) and (2). It is convenient to choose the variable for Laurent expansion as $ \chi (x,t) $ \cite{Conte}, \cite{C&M}, \cite{Kudryashov}, which is concerned with the singular manifold variable $ \xi (x,t) $ \cite{WTC} as
\be
\chi = \left(  \frac{\xi_{x}}{\xi - \xi_{0}} - \frac{\xi_{xx}}{2\xi_{x}}\right)^{-1} ,\,\,\,  \xi_{x} \neq 0
\ee
and satisfies the equations \cite{Conte}, \cite{C&M}
\be \left\{
\begin{aligned}
\chi_{x} & = 1 + \frac{S}{2} \chi^{2} \\ 
\chi_{t} & = - C + C_{x} \chi - \frac{1}{2} (CS + C_{xx}) \chi^{2}, 
\end{aligned}
\right.
\ee
where
\be
S = \frac{\xi_{xxx}}{\xi_{x}} - \frac{3}{2}\left(\frac{\xi_{xx}}{\xi_{x}} \right)^{2}, \,\,\, C = - \frac{\xi_{t}}{\xi_{x}}.
\ee
The substitution of 
\be
u = \frac{u_{0}}{\chi^{p}}, \,\, v = \frac{v_{0}}{\chi^{q}} 
\ee
in the leading order terms of (1) gives $ p = 2 $, $ q = 0 $, $ u_{0} = 2 \alpha \sigma $ and $ v_{0} = -2 $. The Fuchs indices (resonances) are $ -1; 0; 2; 3 $, that leads to the Laurent expansions
\be \left\{
\begin{aligned}
u & = \frac{2 \alpha \sigma}{\chi^{2}} + \frac{u_{-1}}{\chi} + \tilde{u_{0}} + u_{1} \chi + ...\\ 
v & = - 2\ln \chi + \tilde{v_{0}} + v_{1} \chi + v_{2} \chi^{2} + v_{3} \chi^{3} + ... 
\end{aligned}
\right.
\ee
In order to (1) has the Painlev\'e property it is needed that the three coefficients corresponding to the Fuchs indices should be arbitrary functions. However, this is not takes place. We obtain that $ \beta =0 $ and  $ \tilde{v_{0}} $ and $ C $ should satisfy following two equations:
\be \left\{
\begin{aligned}
\frac{\alpha - 2}{\alpha}\,\, C^{2} - (\tilde{v_{0}}_{t} + C \tilde{v_{0}}_{x}) & = 0 \\ 
C_{t} + C\, C_{x} & = 0
\end{aligned}
\right.
\ee
The second equation in (8) is the Hopf one, and its general solution has the form \cite{P&Z PDE}: $ x = C\,t + F(C)$, where $ F(C) $ is an arbitrary function. 

One can suppose that if the system (1) has some integrable reduction to the system of ordinary differential equations (ODE), i.e. it has an exact solution in terms of a certain variable $ y=y(x, t) $, this reduction should be the same one that converts the eqs.(8) to identities. It is easily seen that if we do not consider the stationary case ($ \xi_{t} = \tilde{v_{0}}_{t} \equiv 0 $), we need the requirement $ \alpha = 2 $. In the travelling wave variable $ y = x - ct $, $ c = const $ the Hopf equation vanish identically. This agree with the reduced system (1) 
\be \left\{
\begin{aligned}
u_{y}+cu-u v_{y}+ \lambda & =0 \\ 
\alpha v_{yy}+cv_{y}+ \sigma u & =0,\\
\end{aligned}
\right. 
\ee
$ u=u(y) $, $ v=v(y) $, $ \lambda $ is the integration constant, possesses the Painlev\'e property only if $ \alpha = 2 $. The exact solution of this system has the form \cite{MSh}:
\ba
v & = & - \ln \left[  e^{-\frac{cy}{2}}\,\,A^{2}\,\left( I_{\nu}(\frac{\kappa}{|c|}\,\,e^{-\frac{cy}{2}})\,+\,B\,K_{\nu}(\frac{\kappa}{|c|}\,\,e^{-\frac{cy}{2}}) \right)^{2} \right] \\ \nonumber
 u & = & - \sigma \left( (v_{y})^{2} -  \kappa^{2}\,e^{-cy} + \dfrac{\lambda}{c} \right) , \,\,\,\, \text{where $ \nu^{2}=\dfrac{1}{4}-\dfrac{\lambda}{c^{3}} $}, \\ \nonumber
\ea
$  \kappa >0 $, $ A $ and $ B $ are arbitrary constants.
The functions $ I_{\nu} $ and $ K_{\nu} $ are the Infeld's and Macdonald's functions respectively. 

The analysis of this solution demonstrates that it does not satisfy the requirement $ u \geq 0 $ and $ v \geq 0 $ in all domain of definition. However it seems interesting. In the case of $ \nu = n+\dfrac{1}{2} $ and $ B > 0 $ we obtain soliton-like solutions for $ e^{v(y)} $, in particular, for $ n = 0 $, $ B = \frac{2 + \pi}{2 \pi} $ in terms of $ e^{-\frac{cy}{2}} $ its form coincides with the well-known Korteweg-de Vries soliton
\ba
e^{v(\,e^{-\frac{cy}{2}})} = \frac{\kappa}{C^{2}|c|}\, sech^{2}\left( \frac{\kappa}{|c|}\,\,e^{-\frac{cy}{2}} + \dfrac{1}{2} \ln \dfrac{2}{\pi}\right). 
\ea
For  $ \nu = \dfrac{1}{2} $ and arbitrary $ B $ the function $ u(y) $ is
\be
u(y)  = \dfrac{ \sigma (\pi \, B - 1 )\,\kappa^{2}\,e^{-cy} }{\left( \sinh (\dfrac {\kappa}{|c|}\,\,e^{-\frac{cy}{2}})  +  \dfrac{\pi}{2} \, B\, e^{- \dfrac{\kappa}{|c|}\,\,e^{-\frac{cy}{2}}}  \right)^{2}  }  
\ee
and one can see that for the case of increase of a chemical substance 
( $ \sigma = 1 $ ) the  cells density $ u(y) \geq 0 $ for $ B \geq \dfrac{1}{\pi} $, and for its decrease ($ \sigma = - 1 $) $ u(y) \geq 0 $ for $ B \leq \dfrac{1}{\pi} $. It is obviously also that for $ B > 0 $ $ u(y) $ is the solitary continuous solution vanishing for $ y \rightarrow \pm \infty $, whereas for $ B < 0 $ $ u(y) $ has a point of discontinuity. One can say that when $ B < 0 $ we obtain "blow up" solution in the sense that it goes to infinity for finite $ y $, and this is true for different $ \nu $. 

The others reductions of (1) are not integrable. Besides one-parametric group of translation $ (x, t, v)\rightarrow (x+c\epsilon, t+\epsilon, v ) $ the system (1) has two one-parametric symmetry group of scaling transformation and of shift of $ v $. The invariants variables are $ y = \frac{x}{\sqrt{t}} $ and $ w_{i} $, $ i= 1, 2 $ which satisfy the equation $ (x \partial_{x} + 2t \partial_{t} - 2u \partial_{u} + a \partial_{v}) w_{i} = 0 $, $ a $ is arbitrary constant. One may show that the system of ODEs $ E_{i} (y, {w_{j}}_{yy}, ...) $, $ i, j = 1, 2 $ for different $ w_{i} $ does not possess the Painlev\'e property. 

Now let us rewrite the system (2) in terms of function $ \upsilon (x, t) = \ln v(x, t) $:
\be 
\left\{
\begin{aligned}
u_{t}- u_{xx} + \eta (u \upsilon_{x})_{x} & = 0 \\ 
\upsilon_{t}- \alpha \upsilon_{xx} - \alpha  (\upsilon_{x})^{2} - \beta  + \sigma u & = 0,\\ 
\end{aligned}
\right. 
\tag{\ref{eq:2}$'$}
\ee
The substitution (6) for $ u $ and $ \upsilon $ gives $ p = 2 $, $ q = 0 $, $ u_{0} = \dfrac{2 (2 + \eta ) \alpha \sigma  }{\eta ^{2}} $ and $ \upsilon_{0} = - \dfrac{2}{\eta} $. The Fuchs indices are $ -1; 0; 3 $ and $ r_{4} =   \dfrac{2 (2 + \eta ) }{\eta } $. The requirement $ r_{4} \in \mathbb N, r_{4} \neq 3 $ puts on the restriction on the possible values of $ \eta $. So, for $ r_{4} =  1 $, $ \eta = -4 $ ("negative taxis"), and for $ r_{4} >  1 $, $ \eta > 0 $ ("positive taxis"). Further analysis demonstrates that just as (1), the system (2$'$) has not the Painlev\'e property. For $ \eta = -4 $ we obtain that $ \alpha  = 2  $ and the coefficient $ u_{-1} $ in Laurent expansions should satisfy the equation
\be 
{u_{-1}}_{t}  + C \, {u_{-1}}_{x} = 0.\\ 
\ee
For $ \eta > 0 $  the Hopf equation like (8) has to be satisfied
\be
C_{t} + C \, C_{x}  = 0.
\ee
Therefore we consider (2$'$) as ODE system for functions $ u (y) $ and $ \upsilon (y) $ with $ y = x - ct $:
\be \left\{
\begin{aligned}
u_{y} + c u -  \eta u \upsilon_{y}+ \lambda & =0 \\ 
\alpha \upsilon_{yy} + \alpha (\upsilon_{y})^{2} + c \upsilon_{y} + \beta - \sigma u & =0\\
\end{aligned}
\right. 
\ee
and examine if it has the Painlev\'e property. Substituting (6) for $ u(y) $ and $ \upsilon (y) $ with $ \xi = y - y_{0} $ into (16) gives a similar result: $ p = 2 $, $ q = 0 $, $ u_{0} = \dfrac{2 (2 + \eta ) \alpha \sigma  }{\eta ^{2}} $ and $ \upsilon_{0} = - \dfrac{2}{\eta} $; the Fuchs indices are $ -1; 0 $ and $ r_{3} =  \dfrac{2 (2 + \eta ) }{\eta } $. As a result we have that when $ \alpha = 2 $ for some $ \lambda $ the equation for $ \upsilon (y) $
\be
2\upsilon_{yyy} + 3c \upsilon _{yy} + (c^{2} - \eta \beta ) \upsilon_{y} + 2(2 - \eta )\upsilon_{y} \upsilon_{yy} + 2(2 - \eta ) (\upsilon_{y})^{2} -2\eta (\upsilon_{y})^{3} + c\beta + \sigma \lambda =0
\ee
can be linearized. So, for $ \eta = -4 $ ($ r_{3} =  1 $) the replacement $ \upsilon = \dfrac{1}{2} \ln F $ leads to the third order linear homogeneous  equation with arbitrary $ \lambda $:
\be
2F_{yyy} + 3c F_{yy} + (c^{2} +4\beta ) F_{y} + (c\beta + \sigma \lambda) F =0
\ee
whose solution is well known. For $ \eta > 0 $ ($ r_{3} \geq 3 $) and with $ \lambda = - \sigma c \beta \left( 1 + \dfrac{\eta}{2} \right)  $ we obtain that (16) is equivalent to the following linear equation for $ F $:
\be
F_{y} + c F = 0, \,\,\,\text{where} \,\, \,\,\,    
F(y) = e^{2 \upsilon} \left( 2 \upsilon_{yy} + c \upsilon_{y} - \eta (\upsilon_{y})^{2} - \dfrac{\eta \beta} {2} \right) , 
\ee
that gives the equation for $ \upsilon (y) $
\be
2 \upsilon_{yy} + c \upsilon_{y} - \eta (\upsilon_{y})^{2} - \dfrac{\eta \beta} {2} = C_{1} e^{-2 \upsilon -cy}, 
\ee
$ C_{1} = const $.
However we can obtain an analytical solution for $ \upsilon (y) $ only in several cases. To verify this, we rewrite (19) for the function   $ \Psi (\varsigma) = e^{- \frac{\eta}{2} \upsilon} $ and $ \varsigma = e^{-\frac{cy}{2}} $:
\be
\varsigma ^{2} \Psi_{\varsigma \varsigma}  + \dfrac{\eta^{2} \beta}{2c^{2}} \,\Psi + \dfrac{\eta C_{1}}{c^{2}} \, \varsigma ^{2} \, \Psi ^{\frac{4}{\eta} + 1} = 0.
\ee
A solvable equations of this type are presented in \cite{P&Z ODE}. One can see that for arbitrary $ \beta $ and $ C_{1} $ we have three cases with $ \eta = -4; -2; -1 $, that corresponds to repulsive chemotaxis anf will be considered in future. Thus for "positive taxis", that is for $ \eta > 0 $, we can integrate this equation if we take $ \beta $ or (and) $ C_{1} $ equal to zero. The analysis for $ \beta = 0 $ and the existence of solution is performed in \cite{Wang}. We can write the formal solution for the initial function $ v(y) = e ^{\upsilon (y)} $ when $ \alpha = 2 $  in quadratures:
\be
\int \dfrac{dv}{\sqrt{\sigma |C|\,v^{\eta + 2} - C_{1}}} = \pm \dfrac{2}{|c|\,\sqrt{\eta + 2}}\,\,e^{- \frac{cy}{2}} + \tilde{C}.
\ee
For $ \eta = 1; 2; 4 $, or $ r_{3} = 6; 4; 3 $, this integral can be  expressed in terms of elliptic functions \cite{PBM}.  
 
When $ C_{1} = 0 $, that means $ F(y) =0 $, the equation (20) becomes Euler one, and its solution has three forms according to $ \dfrac{\eta^{2} \beta}{2c^{2}} \lesseqqgtr \dfrac{1}{4} $ \cite{P&Z ODE}. Since $ v(y) $ should be nonnegative and bounded function in $ y \rightarrow \pm \infty $, the appropriate solution is 
\ba
v & = & A \, e^{\frac{c}{2\eta}\,y} \, sech^{\frac{2}{\eta}} \left(  \frac{cb}{2}\,y + b_{0} \right)\\ \nonumber
 u & = &  -\sigma \dfrac{c^{2} b\, (2 + \eta)}{2 \eta ^{2}}  \left\lbrace   \sinh  (cb y + 2 b_{0}) + 2b  \right\rbrace  \, sech^{2} \left(  \frac{cb}{2}\,y + b_{0} \right) + \sigma \dfrac{c^{2}}{\eta^{2}} \left(  1 + \dfrac{\eta}{2}\right),   \\ \nonumber
\ea
where $ A > 0 $, $  b_{0} $ are arbitrary constants, $ b^{2}  = \dfrac{1}{4} + \dfrac{\eta^{2} |\beta|}{2c^{2}} $ and $ \beta < 0 $. For $v(y)$ this is a solitary positive solution, whereas $u(y)$ is alternating function. One can see that $ \sigma u $ has a negative minimum and at least one point where $ u = 0 $. 

The others reductions of (2$'$) as well as (1) are not integrable. So for $ \eta > 0 $ the system of ODEs for $ w_{i} $, $ i= 1, 2 $ which satisfy the equation $ (x \partial_{x} + 2t \partial_{t} - 2u \partial_{u} + 2 \beta t \partial_{v}) w_{i} = 0 $ in terms of $ y = \frac{x}{\sqrt{t}} $  does not have the Painlev\'e property.

\end{document}